\documentclass[12pt,thmsa]{article}
\usepackage{amssymb}

\input{tcilatex}
\begin{document}

\author{Karl-Georg Schlesinger \\
%EndAName
Institute for Theoretical Physics\\
University of Vienna\\
Boltzmanngasse 5\\
A-1090 Vienna, Austria\\
e-mail: kgschles@esi.ac.at}
\title{Does the fivebrane have a nonclassical BV-structure?}
\date{}
\maketitle

\begin{abstract}
The fivebrane in M-theory comes equipped with a higher order gauge field
which should have a formulation in terms of a 2-gerbe on the fivebrane. One
can pose the question if the BV-quantization scheme for such a higher order
gauge theory should differ from the usual BV-algebra structure. We give an
algebraic argument that this should, indeed, be the case and a fourth order
equation should appear as Master equation, in this case. We also discover a
second order term in this equation which seems to indicate that deformation
theory (i.e. solving the Master equation) in this case involves a nonlinear
algebraic theory which goes beyond complexes and cohomology.
\end{abstract}

\section{Introduction}

The fivebrane in M-theory comes equipped with a higher order gauge field
which - due to the coupling to the 4-form G-flux of eleven dimensional
supergravity - should be given by a 2-gerbe on the fivebrane. In the case of
usual gauge theories, given by principal bundles on manifolds, the
BV-quantization scheme can, roughly speaking, be seen as considering the
complex and cohomology - with values in the space of functions on
connections - of the Lie algebra of the gauge group. It is therefore of no
surprise that the general structure of BRST-cohomology resembles Lie algebra
cohomology (or also group cohomology): One has a differential, a
Gerstenhaber bracket (called the BV-antibracket), and a Maurer-Cartan
equation (called the Master equation, in this context). One can now pose the
question if this general structure might be different for the quantization
of the higher gauge theory on the fivebrane. Of course, quantization of the
degrees of freedom of the fivebrane is a deep and completely unsolved issue.
We do certainly not claim to present any solution to this important problem,
here. What we do is studying, only, a deformation question for a general
algebraic structure which should be implied by a 2-gerbe formulation of the
higher gauge field. Then we calculate the analogue of the Maurer-Cartan
equation for the deformation theory of this algebraic structure. We find a
fourth order equation which might be seen as a hint that a new BV-structure
could emerge in the M5-brane case. We also find a new second order term
which seems to indicate that the deformation theory linked to this higher
order Master equation is not determined by cohomology alone but involves a
nonlinear algebraic level beyond complexes and cohomology.

In section 2, we study the algebraic deformation problem in the context of
monoidal bicategories. In section 3, we show that the general algebraic
structure which we find in section 2 is stable under passing to the more
general setting of the deformation theory of enriched categories. Section 4
contains some concluding remarks.

For a different approach to the deformation theory of monoidal bicategories
- which is, especially, structurally not stable under the generalization to
the setting of enriched categories and considers a slightly different
``gauge freedom'' for monoidal bicategories - we would like to refer to \cite
{Elg 2002}, \cite{Elg 2003}.

\bigskip

\section{Deformation theory of monoidal bicategories}

Suppose an M5-brane with world-volume $X$ together with its higher order
gauge theory, in the form of a 2-gerbe on $X$, is given. Whatever the
details of a precise formulation of this 2-gerbe might be, can one say
something about the general algebraic structure which should appear for its
local gauge transformations? Since it seems to be clear that the 2-gerbe
should be related to a kind of principal bundle with bicategories as fibers,
one has to expect that the gauge symmetry is given by a monoidal bicategory
(see \cite{GPS} for the technical definition) instead of a group. Now, as
indicated in the introduction, it is basically the fact that we have a gauge 
\textit{group} for a usual gauge theory with principal bundle which is
responsible for the structure of a BV-algebra appearing for the
BV-quantization scheme, i.e. already the deformation theory of groups leads
one to the correct \textit{general} algebraic structure for the
BV-quantization scheme. It is therefore tempting to speculate that
quantization of the higher gauge theory on the M5-brane should - concerning
the general algebraic structure - be related to the deformation problem for
monoidal bicategories. We will, in this paper, study the question if
BV-quantization of a 2-gerbe gauge theory might involve a nonclassical
Master equation from precisely this point of view.

What are the relevant structures which one can deform, in this case? A
bicategory consists of objects, 1-morphisms, and 2-morphisms and two types
of compositions. For a monoidal bicategory there is, in addition, a third
product structure given by $\otimes $ as a functor of bicategories. In a
bicategory not only the associativity of $\otimes $ can be weak (as in a
monoidal category) but also associativity of the composition of 1-morphisms
needs only to hold up to 2-isomorphisms. So, we have two types of
associators involved. In addition, the exchange rule for $\otimes $ with a
product $\bullet $ 
\begin{equation}
\left( a\otimes b\right) \bullet \left( c\otimes d\right) =\left( a\bullet
c\right) \otimes \left( b\bullet d\right)  \label{1}
\end{equation}
needs only to hold in the weak form for the exchange of $\otimes $ with the
composition of 1-morphisms (for the details of the weak rules, see, again 
\cite{GPS}). So, we have a new kind of ``weak exchange morphism'' appearing
in addition to the two associators for monoidal bicategories.

Suppose for a moment that all structures would be strict, i.e. we would have
strict associativity for all three products and the strict exchange rule (%
\ref{1}) would hold for all pairs of products. The deformation theory of a
strict monoidal category was studied in \cite{GS} and it was discovered
there that the Maurer-Cartan equation is replaced by a system of three
coupled differential equations, consisting of two Maurer-Cartan equations
(for the associativity of the two products) plus a constraint (for the
exchange rule). The structure of the deformation complex receives an
additional ingredient beyond the Gerstenhaber bracket which resembles a
curvature tensor on the complex. So, one gets a kind of nonlinear complex
for the deformation theory.

It is straightforward to see that this structure generalizes to the case of
a strict monoidal bicategory (also called a monoidal 2-category in the
literature). One gets a coupled system of three Maurer-Cartan equations plus
constraints for all the exchange rules, in this case. The general structure
of the deformation complex remains very much of the same type as found for
strict monoidal categories. For monoidal categories, one can exclude the
case of a nontrivial associator for $\otimes $ and the corresponding
deformation theory for the associator because it can always be gauged away
by Mac Lane's coherence theorem, i.e. the deformation theory of an
associator should contribute trivially to the general deformation theory of
a monoidal category. In physical terms one has learned to see the existence
of a coherence theorem for the weak structure which allows to gauge
transform to the strict one as saying that the additional data of the weak
structure arise from BRST-exact terms, only (compare e.g. to the $A_\infty $%
-structures appearing in topological string theory).

For the case of monoidal bicategories it is known that a coherence theorem
in this sense does not exist but one has only equivalence of a general
monoidal bicategory to a so called semistrict monoidal bicategory (instead
of equivalence to a strict monoidal bicategory, see \cite{GPS}). Roughly
speaking, this means that we can always gauge away the two associators
involved in the definition of a monoidal bicategory but we can, in general,
not gauge away the ``weak exchange morphism''. It turns out that it suffices
to have a ``weak exchange morphism'' for (\ref{1}) with $b$ and $c$
identities which is what is called a cubic functor in the literature (see 
\cite{GPS}). We will write this down in more detail, below.

So, for monoidal bicategories we have to include the deformation theory of
the cubic functor, in addition to the deformation theory of a strict
monoidal bicategory because this should lead to non-BRST-exact contributions
which should therefore be relevant to the structure on observables. Since
the deformation theory of a strict monoidal bicategory would basically mean
redoing \cite{GS}, we exclude this case, here, and restrict to the cubic
functor, alone. So, the deformation problem which we consider is the problem
of deforming a semistrict monoidal bicategory into another semistrict
monoidal bicategory such that $\otimes $ and all compositions of morphism
remain fixed and the cubic functor is deformed into a new one.

\bigskip

\begin{remark}
Monoidal bicategories can be seen as tricategories with one object. So, the
deformation problem considered, here, is precisely the deformation problem
which was suggested in \cite{Sch} to be linked to the study of the
universality question of noncommutative field theories (passing to the case
of more than one object does not make a difference for the deformation
theory). A semistrict tricategory is sometimes also called a \textbf{Gray}%
-category (category enriched over the monoidal category \textbf{Gray} of
2-categories with the \textbf{Gray}-tensor product - see \cite{GPS} - which
induces the nontrivial cubic functors). So, a semistrict monoidal bicategory
is a \textbf{Gray}-category with one object.
\end{remark}

\bigskip

Let us now write down the data of a cubic functor in a little bit more
detail (for the full technical definition, see \cite{GPS} or the nice review
included in \cite{Lau}). Let 
\[
f:A\rightarrow A^{\prime } 
\]
and 
\[
g:B\rightarrow B^{\prime } 
\]
be 1-morphisms in the semistrict monoidal bicategory $\mathcal{C}$. Then
there should always be given a 2-isomorphism $K_{f,g}$ 
\[
K_{f,g}:\left( f\otimes 1_{B^{\prime }}\right) \left( 1_A\otimes g\right)
\rightarrow \left( 1_{A^{\prime }}\otimes g\right) \left( f\otimes
1_B\right) 
\]
where juxtaposition means composition of 1-morphisms. The whole family $%
K_{f,g}$ gives the central part of the data of a cubic functor. From the
conditions imposed on $K_{f,g}$, we will keep only one for the deformation
theory. The reason is that Hochschild or also Hopf algebra cohomology shows
that only the constraints on product like structures - like associativity or
coassociativity - have to be imposed while e.g. constraints on unital
elements come along, automatically. We will, heuristically, proceed as if
this would also be true for the cubic functor structure. But the analogous
theorems are an open problem for future work, in this case.

Denoting horizontal composition of 2-morphisms by $\circ $ and vertical
composition by juxtaposition, the constraint reads as 
\begin{eqnarray}
&&\left( K_{f^{\prime },g^{\prime }}K_{f^{\prime },g}\right) \circ \left(
K_{f,g^{\prime }}K_{f,g}\right)  \label{2} \\
&=&\left( K_{f^{\prime },g^{\prime }}\circ K_{f,g^{\prime }}\right) \left(
K_{f^{\prime },g}\circ K_{f,g}\right)  \nonumber \\
&=&K_{ff^{\prime },gg^{\prime }}  \nonumber
\end{eqnarray}
We will now assume that $\mathcal{C}$ is a $\Bbb{C}$- or $\Bbb{R}$-linear,
semistrict monoidal bicategory, i.e. all morphism spaces are ($\Bbb{C}$- or $%
\Bbb{R}$-) vector spaces and both compositions of morphisms, as well, as $%
\otimes $ are assumed to be bilinear. We proceed in the same way, then, as
e.g. in the case of the usual deformation theory of an associative algebra.
For simplicity, we will always neglect the fact that actually one passes to
a category of formal power series with coefficients in $\mathcal{C}$ in the
deformation theory.

We replace the cubic functor $K_{f,g}$ by a new cubic functor $\widetilde{%
K_{f,g}}$ with 
\begin{equation}
\widetilde{K_{f,g}}=K_{f,g}+\Psi _{f,g}  \label{3}
\end{equation}
where $\Psi _{f,g}$ is a collection of 2-morphisms in $\mathcal{C}$ (observe
that we assume both compositions in $\mathcal{C}$, as well, as $\otimes $ to
be fixed in the process of the deformation). Next, we want to calculate the
analogue of the Maurer-Cartan equation for $\Psi _{f,g}$.

By inserting (\ref{3}) into (\ref{2}) written for $\widetilde{K_{f,g}}$, we
get in 0-th order the original equation (\ref{2})\ for $K_{f,g}$. In first
order, we get for the left hand side 
\begin{eqnarray*}
&&\left( K_{f^{\prime },g^{\prime }}K_{f^{\prime },g}\right) \circ \left(
K_{f,g^{\prime }}\Psi _{f,g}\right) +\left( K_{f^{\prime },g^{\prime
}}K_{f^{\prime },g}\right) \circ \left( \Psi _{f,g^{\prime }}K_{f,g}\right)
\\
&&+\left( K_{f^{\prime },g^{\prime }}\Psi _{f^{\prime },g}\right) \circ
\left( K_{f,g^{\prime }}K_{f,g}\right) +\left( \Psi _{f^{\prime },g^{\prime
}}K_{f^{\prime },g}\right) \circ \left( K_{f,g^{\prime }}K_{f,g}\right)
\end{eqnarray*}
and for the right hand side 
\begin{eqnarray*}
&&\left( K_{f^{\prime },g^{\prime }}\circ K_{f,g^{\prime }}\right) \left(
K_{f^{\prime },g}\circ \Psi _{f,g}\right) +\left( K_{f^{\prime },g^{\prime
}}\circ K_{f,g^{\prime }}\right) \left( \Psi _{f^{\prime },g}\circ
K_{f,g}\right) \\
&&+\left( K_{f^{\prime },g^{\prime }}\circ \Psi _{f,g^{\prime }}\right)
\left( K_{f^{\prime },g}\circ K_{f,g}\right) +\left( \Psi _{f^{\prime
},g^{\prime }}\circ K_{f,g^{\prime }}\right) \left( K_{f^{\prime },g}\circ
K_{f,g}\right)
\end{eqnarray*}
where both have to be equal to 
\[
\Psi _{ff^{\prime },gg^{\prime }} 
\]
For the second order terms, we calculate for the left hand side 
\begin{eqnarray*}
&&\left( K_{f^{\prime },g^{\prime }}K_{f^{\prime },g}\right) \circ \left(
\Psi _{f,g^{\prime }}\Psi _{f,g}\right) +\left( K_{f^{\prime },g^{\prime
}}\Psi _{f^{\prime },g}\right) \circ \left( K_{f,g^{\prime }}\Psi
_{f,g}\right) \\
&&+\left( K_{f^{\prime },g^{\prime }}\Psi _{f^{\prime },g}\right) \circ
\left( \Psi _{f,g^{\prime }}K_{f,g}\right) +\left( \Psi _{f^{\prime
},g^{\prime }}K_{f^{\prime },g}\right) \circ \left( K_{f,g^{\prime }}\Psi
_{f,g}\right) \\
&&+\left( \Psi _{f^{\prime },g^{\prime }}K_{f^{\prime },g}\right) \circ
\left( \Psi _{f,g^{\prime }}K_{f,g}\right) +\left( \Psi _{f^{\prime
},g^{\prime }}\Psi _{f^{\prime },g}\right) \circ \left( K_{f,g^{\prime
}}K_{f,g}\right)
\end{eqnarray*}
and for the right hand side 
\begin{eqnarray*}
&&\left( K_{f^{\prime },g^{\prime }}\circ K_{f,g^{\prime }}\right) \left(
\Psi _{f^{\prime },g}\circ \Psi _{f,g}\right) +\left( K_{f^{\prime
},g^{\prime }}\circ \Psi _{f,g^{\prime }}\right) \left( K_{f^{\prime
},g}\circ \Psi _{f,g}\right) \\
&&+\left( K_{f^{\prime },g^{\prime }}\circ \Psi _{f,g^{\prime }}\right)
\left( \Psi _{f^{\prime },g}\circ K_{f,g}\right) +\left( \Psi _{f^{\prime
},g^{\prime }}\circ K_{f,g^{\prime }}\right) \left( K_{f^{\prime },g}\circ
\Psi _{f,g}\right) \\
&&+\left( \Psi _{f^{\prime },g^{\prime }}\circ K_{f,g^{\prime }}\right)
\left( \Psi _{f^{\prime },g}\circ K_{f,g}\right) +\left( \Psi _{f^{\prime
},g^{\prime }}\circ \Psi _{f,g^{\prime }}\right) \left( K_{f^{\prime
},g}\circ K_{f,g}\right)
\end{eqnarray*}
As third order terms, we get 
\begin{eqnarray*}
&&\left( K_{f^{\prime },g^{\prime }}\Psi _{f^{\prime },g}\right) \circ
\left( \Psi _{f,g^{\prime }}\Psi _{f,g}\right) +\left( \Psi _{f^{\prime
},g^{\prime }}K_{f^{\prime },g}\right) \circ \left( \Psi _{f,g^{\prime
}}\Psi _{f,g}\right) \\
&&+\left( \Psi _{f^{\prime },g^{\prime }}\Psi _{f^{\prime },g}\right) \circ
\left( K_{f,g^{\prime }}\Psi _{f,g}\right) +\left( \Psi _{f^{\prime
},g^{\prime }}\Psi _{f^{\prime },g}\right) \circ \left( \Psi _{f,g^{\prime
}}K_{f,g}\right)
\end{eqnarray*}
respectively 
\begin{eqnarray*}
&&\left( K_{f^{\prime },g^{\prime }}\circ \Psi _{f,g^{\prime }}\right)
\left( \Psi _{f^{\prime },g}\circ \Psi _{f,g}\right) +\left( \Psi
_{f^{\prime },g^{\prime }}\circ K_{f,g^{\prime }}\right) \left( \Psi
_{f^{\prime },g}\circ \Psi _{f,g}\right) \\
&&+\left( \Psi _{f^{\prime },g^{\prime }}\circ \Psi _{f,g^{\prime }}\right)
\left( K_{f^{\prime },g}\circ \Psi _{f,g}\right) +\left( \Psi _{f^{\prime
},g^{\prime }}\circ \Psi _{f,g^{\prime }}\right) \left( \Psi _{f^{\prime
},g}\circ K_{f,g}\right)
\end{eqnarray*}
Finally, the fourth order terms are 
\[
\left( \Psi _{f^{\prime },g^{\prime }}\Psi _{f^{\prime },g}\right) \circ
\left( \Psi _{f,g^{\prime }}\Psi _{f,g}\right) 
\]
for the left hand side and 
\[
\left( \Psi _{f^{\prime },g^{\prime }}\circ \Psi _{f,g^{\prime }}\right)
\left( \Psi _{f^{\prime },g}\circ \Psi _{f,g}\right) 
\]
for the right hand side, respectively.

Let us now discuss the structural properties of these terms. Consider $%
K_{.,.}$ as a map of two variables from 1-morphisms in $\mathcal{C}$ to
2-morphisms in $\mathcal{C}$. Since 
\[
K_{f,g}:\left( f\otimes 1_{B^{\prime }}\right) \left( 1_A\otimes g\right)
\rightarrow \left( 1_{A^{\prime }}\otimes g\right) \left( f\otimes
1_B\right) 
\]
$K_{f,g}$ measures the failure of the square shaped diagram, defined by the
1-morphisms $f$ and $g$ as given above, to commute. We can now assemble four
such diagrams as two times two into a larger square shaped diagram, as
required for equation (\ref{2}). Obviously, the failure of this larger
diagram to commute is then measured by a map of four variables from
1-morphisms in $\mathcal{C}$ to 2-morphisms in $\mathcal{C}$. Equation (\ref
{2})\ states that in the case of $K_{f,g}$ no new family of maps arises in
this way but the four variable map is independent of the two ways to compose
in the big two times two diagram and is given by $K_{ff^{\prime },gg^{\prime
}}$. For a general map of two variables from 1-morphisms in $\mathcal{C}$ to
2-morphisms in $\mathcal{C}$ this need not be true and for such a general
map $\Phi _{f,g}$, we define $d_K\Phi _{f,g}$ as the four variable map given
by 
\begin{eqnarray*}
&&d_K\Phi _{f,g} \\
&=&\left( K_{f^{\prime },g^{\prime }}K_{f^{\prime },g}\right) \circ \left(
K_{f,g^{\prime }}\Phi _{f,g}\right) +\left( K_{f^{\prime },g^{\prime
}}K_{f^{\prime },g}\right) \circ \left( \Phi _{f,g^{\prime }}K_{f,g}\right)
\\
&&+\left( K_{f^{\prime },g^{\prime }}\Phi _{f^{\prime },g}\right) \circ
\left( K_{f,g^{\prime }}K_{f,g}\right) +\left( \Phi _{f^{\prime },g^{\prime
}}K_{f^{\prime },g}\right) \circ \left( K_{f,g^{\prime }}K_{f,g}\right) \\
&&-\left( K_{f^{\prime },g^{\prime }}\circ K_{f,g^{\prime }}\right) \left(
K_{f^{\prime },g}\circ \Phi _{f,g}\right) -\left( K_{f^{\prime },g^{\prime
}}\circ K_{f,g^{\prime }}\right) \left( \Phi _{f^{\prime },g}\circ
K_{f,g}\right) \\
&&-\left( K_{f^{\prime },g^{\prime }}\circ \Phi _{f,g^{\prime }}\right)
\left( K_{f^{\prime },g}\circ K_{f,g}\right) -\left( \Phi _{f^{\prime
},g^{\prime }}\circ K_{f,g^{\prime }}\right) \left( K_{f^{\prime },g}\circ
K_{f,g}\right)
\end{eqnarray*}
Here, the notation $d_K$ indicates the dependence on the cubic functor $K$
(understood as the whole family $K_{f,g}$). The requirement 
\[
d_K\Phi _{f,g}=0 
\]
means, then, that, at least in first order correction, new contributions to
the deficiency of the big square diagram to commute - besides the ones
deriving from the deficiency in commutativity of the four small squares -
vanish. This is precisely what is satisfied in a linear approximation by $%
\Psi _{f,g}$.

Starting from the square shaped diagram and then building the big two times
two diagram out of this, one can proceed in this way. In the next step, one
constructs a ``huge diagram'', consisting of two times two big diagrams.
Obviously, the deficiency of the huge diagram to commute is measured by a
function of eight variables. We can proceed iteratively, then. Starting from
a given diagram and its deficiency function, we can pass to the next larger
diagram and calculate in first order the additional contribution to the
noncommutativity of the larger diagram, besides the contribution arising
form the deficiency function of the four smaller diagrams. In the same way
as above, we define a differential $d_K$ for the deficiency function of the
smaller diagram, in this way. One has a coherence property in this setting:
Building larger and larger two times two diagrams completely reduces to
iterating the first step which leads from the square shaped diagram to the
big square shaped diagram. It follows from this coherence property that 
\begin{equation}
d_K^2=0  \label{4}
\end{equation}
We complete the construction of the deformation complex by including as
0-cochains the real or complex numbers (depending on wether $\mathcal{C}$ is 
$\Bbb{R}$- or $\Bbb{C}$-linear). We interpret a number $\lambda $ as giving
the trivial deficiency function $K_{.,.}$ which is constantly the identity,
rescaled (remember that we are in a linear setting) by the factor $\lambda $%
. The cocycle condition 
\[
d_K\lambda =0 
\]
is trivial, then. Using (\ref{4}), we can pass from the deformation complex
to cohomology.. Observe that the resulting cohomology theory has two exotic
features:

\bigskip

\begin{itemize}
\item  It is a strange exponential cohomology:\ While e.g. in Hochschild
cohomology of associative algebras the $n$-cochains are $n$-variable maps,
here, the $n$-cochains are maps of $2^n$ variables.

\item  As a result of this exponential behavior, the first order
deformations of a cubic functor live in first cohomology while in Hochschild
cohomology the first order deformations live in second cohomology.
\end{itemize}

\bigskip

Observe that we can not expect for an analogue of the result of \cite{Kon
1997} to hold for the deformation theory of a cubic functor: Since we have a
fourth order Master equation, we can not expect deformations to be
determined by first order terms but we expect that we also have to prescribe
a second and a third order term to determine a deformation. Hence, we can
not expect that obstructions to extending beyond first order deformations
are of a purely cohomological nature. We will next try to understand the
higher order terms of the Master equation in more detail.

Let us start with the fourth order terms. Define the difference of the left
and the right hand side of (\ref{2}) as an operator $J\left( K\right) $,
i.e. (\ref{2}) can be rephrased as 
\begin{equation}
J\left( K\right) =0  \label{5}
\end{equation}
We can extend this operator to any two variable map $\Phi $ by just
replacing $K$ by $\Phi $ in the definition. The fourth order terms are just
of the form $J\left( \Psi \right) $, then. The operator $J$ is independent
of the given cubic functor $K$. This is analogous to Hochschild cohomology
where the highest order terms appearing in the Maurer-Cartan equation 
\begin{equation}
d_ma+a\circ _Ga=d_ma+\frac 12\left[ a,a\right] _G=0  \label{6}
\end{equation}
are given by the Gerstenhaber product $\circ _G$ or its odd commutator in
the form of the Gerstenhaber bracket $\left[ ,\right] _G$ and are also
independent of the given algebra product $m$ one wants to deform. Also,
associativity of the original product $m$ is expressed as 
\begin{equation}
\left[ m,m\right] _G=0  \label{7}
\end{equation}
in the Hochschild setting which is analogous to (\ref{5}). The Maurer-Cartan
equation (\ref{6}) says that the deficiency of $a$ to satisfy (\ref{7}) is
measured by the differential $d_m$ applied to $a$. The Master equation for
the deformation of a cubic functor says that the deficiency of $\Psi $ to
satisfy (\ref{5})\ is measured by the differential $d_K$ applied to $\Psi $,
plus the second and third order terms.

The third order terms can be concisely summarized as 
\[
d_\Psi K 
\]
where we define $d_\Psi $ by the formula resulting from the definition of $%
d_K$ after replacing $K$ by $\Psi $. Since $\Psi $ is not a cubic functor,
we have to expect that in general 
\[
d_\Psi ^2\neq 0 
\]
This means that in the same way as $J\left( \Psi \right) $ normally does not
vanish, in general there is a deficiency to $d_\Psi $ being a true
differential. So, besides the symmetry between the 0-th and the highest
order term under exchanging $K$ with $\Psi $ - which is also known from the
Maurer-Cartan equation of Hochschild cohomology - we have discovered another
symmetry of this type for the Master equation for the deformation theory of
a cubic functor, as a symmetry between the first and third order terms.

Finally, we summarize the second order terms as 
\[
\Psi \boxdot _K\Psi 
\]
where $\boxdot _K$ can be extended in the obvious way to a product $\Phi
_1\boxdot _K\Phi _2$ on general two variable maps $\Phi _1$ and $\Phi _2$.
The Master equation for the deformation theory of a cubic functor can then
be written as 
\begin{equation}
d_K\Psi +\Psi \boxdot _K\Psi +d_\Psi K+J\left( \Psi \right) =0  \label{8}
\end{equation}
While the Maurer-Cartan equation says that the deficiency of $a$ to satisfy (%
\ref{7}) is given by $d_ma$, we can interpret equation (\ref{8}) also as
follows: Besides the cubic functor $K$, we have to consider its cohomology
given by cocycles with 
\begin{equation}
d_K\Psi =0  \label{9}
\end{equation}
In general 
\[
J\left( \Psi \right) \neq 0 
\]
and 
\[
d_\Psi K\neq 0 
\]
but the deficiency that at least the weaker condition 
\[
d_\Psi K+J\left( \Psi \right) =0 
\]
would be satisfied is measured by the second order term $\Psi \boxdot _K\Psi 
$ if $\Psi $ is a cocycle, i.e. (\ref{9}) holds. We would like to stress
that the product $\boxdot _K$ is \textit{not} of the type of the
Gerstenhaber product because $\boxdot _K$ depends on $K$, much like the
differential $d_K$ is dependent on $K$. So, besides the cohomological first
order part and a kind of deformed (because of the deficiency of $d_\Psi $ to
square to zero) cohomology given by $d_\Psi $, we have a new structural
ingredient in the Master equation (\ref{8}) in the form of the product $%
\boxdot _K$. So, to determine a solution of the Master equation, a new
nonlinear algebraic level beyond complexes and cohomology is needed.

In the next section, we will study the deformation problem of a cubic
functor from a slightly different perspective. We will see that similar
structural properties for the Master equation arise.

\bigskip

\section{Deformation theory of enriched categories}

As we mentioned in Remark 1, semistrict monoidal bicategories and, more
generally, semistrict tricategories can also be understood as categories
enriched over the monoidal category \textbf{Gray} of 2-categories with the 
\textbf{Gray} tensor product (see \cite{GPS}, see \cite{Kel} for background
on enriched categories). The structure of a cubic functor is induced from
the \textbf{Gray} tensor product. If one uses the usual cartesian product as
tensor product, instead, one gets strict tricategories, i.e. trivial cubic
functors, only. So, we can understand the deformation problem of a cubic
functor also as being related to the deformation problem for the tensor
product of the monoidal category over which the enrichment is taken. We will
explain this in a little bit more detail, now.

Let $\mathcal{V}$ be a monoidal category. A category $\mathcal{C}$ enriched
over $\mathcal{V}$, also called a $\mathcal{V}$-category, is given by the
following data:

\bigskip

\begin{itemize}
\item  A class $Obj\left( \mathcal{C}\right) $ of objects of $\mathcal{C}$.

\item  For any pair of objects $a,b$ of $\mathcal{C}$, we have an object $%
Hom\left( a,b\right) $ in $\mathcal{V}$.

\item  For objects $a,b,c$ of $\mathcal{C}$, there is a composition morphism 
$f_{a,b,c}$ in $\mathcal{V}$ 
\[
f_{a,b,c}:Hom\left( a,b\right) \otimes Hom\left( b,c\right) \rightarrow
Hom\left( a,c\right) 
\]
which is supposed to be associative. Here, $\otimes $ is the tensor product
of $\mathcal{V}$.

\item  In addition, one has straightforward requirements on existence of
units which we do not explicitly give, here, since we will not need them in
the sequel (see e.g. \cite{Kel} for the full details).
\end{itemize}

\bigskip

Deforming $\otimes $ - to pass e.g. from the case of a trivial to a
nontrivial cubic functor - one automatically has to include deformations of
the function system $f_{a,b,c}$ since the domain of definition of $f_{a,b,c}$
is, in general, changed under deformations of the tensor product. We will
consider an even more general situation, here, where we allow for
deformations of the composition of morphisms in $\mathcal{V}$ to become
deformed, too. So, the deformation problem we consider for enriched
categories is, in the case of \textbf{Gray}-categories, even more general
than the one discussed in the previous section. The deformation problem for
an enriched category is then the problem to deform $f_{a,b,c}$, $\otimes $
and the composition $\bullet $ of $\mathcal{V}$ at the same time, i.e we
deform $\mathcal{C}$ into a new enriched category $\widetilde{\mathcal{C}}$
where $\widetilde{\mathcal{C}}$ is enriched over the monoidal category $%
\widetilde{\mathcal{V}}$, resulting from the deformation of $\otimes $ and $%
\bullet $.

We assume, now, that $\mathcal{C}$ and $\mathcal{V}$ are $\Bbb{R}$- or $\Bbb{%
C}$-linear, completely analogous to the corresponding assumption in the
previous section. Let 
\[
\widetilde{f}=f+\alpha 
\]
where we write $f$ for the whole function system $f_{a,b,c}$, 
\[
\widetilde{\otimes }=\otimes +\beta 
\]
and 
\[
\widetilde{\bullet }=\bullet +\gamma 
\]
The first requirement in the deformation theory is then that $\widetilde{%
\bullet }$ and $\widetilde{\otimes }$ constitute a monoidal category, again.
This results in the deformation theory described in \cite{GS}. In addition,
we have the requirement that $\widetilde{f}$ has to be associative. Since
the associativity constraint on $\widetilde{f}$ also involves $\widetilde{%
\otimes }$ and $\widetilde{\bullet }$, this does not result in the usual
Maurer-Cartan equation for the deformation theory of an associative product,
in this case. We will now study the equation which generalizes the
Maurer-Cartan equation, in this case (we will call this the Master equation
for the deformation theory of an enriched category but remember that this
Master equation has to be coupled to the coupled system of three
differential equations of \cite{GS} as a fourth equation, to arrive at the
full deformation theory of an enriched category). We will find a fourth
order equation, again.

The associativity constraint for $\widetilde{f}$ is 
\begin{equation}
\widetilde{f}\widetilde{\bullet }\left( \widetilde{f}\widetilde{\otimes }%
1\right) =\widetilde{f}\widetilde{\bullet }\left( 1\widetilde{\otimes }%
\widetilde{f}\right)  \label{10}
\end{equation}
where we have suppressed indices of $\widetilde{f}$ (actually (\ref{10})
refers to four objects $a,b,c,d$ of $\mathcal{C}$, of course). In the same
way, we have not expelled the index for the unit object $1$. Inserting $%
\widetilde{f}$, $\widetilde{\otimes }$, and $\widetilde{\bullet }$ as given
above into (\ref{10}), we get in 0-th order the associativity constraint 
\[
f\bullet \left( f\otimes 1\right) =f\bullet \left( 1\otimes f\right) 
\]
for $f$. Rewriting (\ref{10}) as 
\[
\widetilde{f}\widetilde{\bullet }\left( \widetilde{f}\widetilde{\otimes }%
1\right) -\widetilde{f}\widetilde{\bullet }\left( 1\widetilde{\otimes }%
\widetilde{f}\right) =0 
\]
we get for the first order terms 
\begin{eqnarray*}
&&f\bullet \left( \alpha \otimes 1\right) -f\bullet \left( 1\otimes \alpha
\right) +f\bullet \beta \left( f,1\right) -f\bullet \beta \left( 1,f\right)
\\
&&+\alpha \bullet \left( f\otimes 1\right) -\alpha \bullet \left( 1\otimes
f\right) +\gamma \left( f,f\otimes 1\right) -\gamma \left( f,1\otimes
f\right)
\end{eqnarray*}
For the second order terms, we have 
\begin{eqnarray*}
&&f\bullet \beta \left( \alpha ,1\right) -f\bullet \beta \left( 1,\alpha
\right) +\alpha \bullet \left( \alpha \otimes 1\right) -\alpha \bullet
\left( 1\otimes \alpha \right) \\
&&+\alpha \bullet \beta \left( f,1\right) -\alpha \bullet \beta \left(
1,f\right) +\gamma \left( f,\alpha \otimes 1\right) -\gamma \left(
f,1\otimes \alpha \right) \\
&&+\gamma \left( f,\beta \left( f,1\right) \right) -\gamma \left( f,\beta
\left( 1,f\right) \right) +\gamma \left( \alpha ,f\otimes 1\right) -\gamma
\left( \alpha ,1\otimes f\right)
\end{eqnarray*}
and the third order terms are given by 
\begin{eqnarray*}
&&\alpha \bullet \beta \left( \alpha ,1\right) -\alpha \bullet \beta \left(
1,\alpha \right) +\gamma \left( f,\beta \left( \alpha ,1\right) \right)
-\gamma \left( f,\beta \left( 1,\alpha \right) \right) \\
&&+\gamma \left( \alpha ,\alpha \otimes 1\right) -\gamma \left( \alpha
,1\otimes \alpha \right) +\gamma \left( \alpha ,\beta \left( f,1\right)
\right) -\gamma \left( \alpha ,\beta \left( 1,f\right) \right)
\end{eqnarray*}
Finally, the fourth order terms result in 
\[
\gamma \left( \alpha ,\beta \left( \alpha ,1\right) \right) -\gamma \left(
\alpha ,\beta \left( 1,\alpha \right) \right) 
\]
Let us comment, again, on the structural properties of these terms. We
summarize the first order terms as 
\[
d_{f,\otimes ,\bullet }\left( \alpha ,\beta ,\gamma \right) 
\]
where the notation indicates the dependence of $d_{f,\otimes ,\bullet }$ on $%
f,\otimes $ and $\bullet $. As in the previous section, there is a symmetry
between the first and third order terms and we can formally write the third
order terms as 
\[
d_{\alpha ,\beta ,\gamma }\left( f,\otimes ,\bullet \right) 
\]
Also, there is a symmetry, again, between the 0-th and the fourth order
terms: Define for a function system $\chi $, or more detailed $\chi _{a,b,c}$%
, of two variables 
\[
\chi _{a,b,c}:Hom\left( a,b\right) \otimes _{lin}Hom\left( b,c\right)
\rightarrow Hom\left( a,c\right) 
\]
(where the notation $\otimes _{lin}$ indicates that the left hand side
agrees with $Hom\left( a,b\right) \otimes Hom\left( b,c\right) $ as taken in 
$\mathcal{C}$ only as a linear space over $\Bbb{R}$ or $\Bbb{C}$; remember
in this context that the linear structures of $\mathcal{C}$ and $\mathcal{V}$
are supposed to stay fixed in the process of deformation), and corresponding
function systems $\varsigma $ and $\rho $, generalizing $\otimes $ and $%
\bullet $, the operator $G$ as 
\[
G\left( \chi ,\varsigma ,\rho \right) =\rho \left( \chi ,\varsigma \left(
\chi ,1\right) \right) -\rho \left( \chi ,\varsigma \left( 1,\chi \right)
\right) 
\]
Then the associativity constraint on $f$ is equivalent to 
\begin{equation}
G\left( f,\otimes ,\bullet \right) =0  \label{11}
\end{equation}
and we can summarize the fourth order terms as 
\[
G\left( \alpha ,\beta ,\gamma \right) 
\]
Again, $G$ is independent of $f,\otimes $ and $\bullet $ and the Master
equation gives a deficiency of $\alpha ,\beta ,\gamma $ to satisfy (\ref{11}%
). As in the previous section, the second order terms can be written as a
product \textit{depending} on $f,\otimes $ and $\bullet $ which we denote as 
$\boxdot _{f,\otimes ,\bullet }$, i.e. the second order terms are given by 
\[
\left( \alpha ,\beta ,\gamma \right) \boxdot _{f,\otimes ,\bullet }\left(
\alpha ,\beta ,\gamma \right) 
\]
So, the whole Master equation for the deformation theory of an enriched
category can be written as 
\begin{equation}
d_{f,\otimes ,\bullet }\left( \alpha ,\beta ,\gamma \right) +\left( \alpha
,\beta ,\gamma \right) \boxdot _{f,\otimes ,\bullet }\left( \alpha ,\beta
,\gamma \right) +d_{\alpha ,\beta ,\gamma }\left( f,\otimes ,\bullet \right)
+G\left( \alpha ,\beta ,\gamma \right) =0  \label{12}
\end{equation}
The discussion of the remaining structural properties of (\ref{12}) is
completely analogous, then, to the discussion for the Master equation of the
deformation theory of a cubic functor in the previous section and we will,
therefore, not repeat it, here. Especially, we expect that, as a consequence
of the coherence property for the associativity constraint (\ref{10}) (i.e.
iterated brackets with four and more variables can be rebracketed
automatically as a consequence of (\ref{10})), 
\[
d_{f,\otimes ,\bullet }^2=0 
\]
holds.

In conclusion, the deformation theory of an enriched category is described
by the Master equation (\ref{12}), which is very similar to the Master
equation of the previous section, but with an additional coupling to the
deformation theory of a strict monoidal category as given in \cite{GS}.

The viewpoint starting from enriched categories for the deformation theory,
as taken in this section, might also be of interest in physics besides the
setting of semistrict monoidal bicategories. E.g. also the differential
graded categories and $A_\infty $-categories appearing in topological string
theory can be understood as enriched categories. The deformation theory of
enriched categories, as presented here, would considerably generalize the
Hochschild complex for these structures and one can pose the question what
the even larger moduli spaces that would arise this way should mean for
topological string theory.

\bigskip

\section{Conclusion}

We have studied the deformation theory of a semistrict monoidal bicategory
in this paper and found structural properties and a Master equation which
considerably differ from the usual structure of the BV-quantization scheme.
We found a similar result when studying the deformation problem not as the
deformation problem for the cubic functor of a semistrict monoidal
bicategory but in the more general setting of enriched categories.

For function algebras on quantum groups, taking the case where the
deformation parameter $q$ is a root of unity, several authors have
undertaken to construct a monoidal bicategory from the data (see e.g. \cite
{CF}, \cite{Soi}). Especially, this means that the deformation theory
developed, here, should be applicable in the setting of the $SU(2)$-WZW
model. We plan to study this application in a separate paper.

\bigskip

\textbf{Acknowledgments:} I would like to thank H. Grosse and V. Mathai for
discussions on or related to the material of this paper.


\bigskip

\bigskip

\begin{thebibliography}{Elg 2002}
\bibitem[CF]{CF}  L. Crane, I. B. Frenkel, \textit{Four-dimensional
topological quantum field theory, Hopf categories and the canonical bases},
J. Math. Phys. 35, 5136-5154 (1994).

\bibitem[Elg 2002]{Elg 2002}  J. Elgueta, \textit{Cohomology and deformation
theory of monoidal 2-categories I}, math.QA/0204099v2.

\bibitem[Elg 2003]{Elg 2003}  J. Elgueta, \textit{2-cosemisimplicial objects
in a 2-category, permutohedra and deformations of pseudofunctors},
math.QA/0309066v2.

\bibitem[Kel]{Kel}  G. M. Kelly, \textit{Basic concepts of enriched category
theory}, London Math. Soc. Lecture Notes Series 64, Cambridge University
Press, Cambridge 1982.

\bibitem[Kon]{Kon 1997}  M. Kontsevich, \textit{Deformation quantization of
Poisson manifolds I}, math/9709180.

\bibitem[KS]{Soi}  D. Kazhdan, Y. Soibelman, \textit{Representations of
algebras of functions on quantum groups at root of unity, 2-categories and
Zamolodchikov tetrahedra equation}, The Gelfand Mathematical Seminar
(1990-1992), Birkh\"{a}user, Basel (1993), 151-163.

\bibitem[Lau]{Lau}  A. Lauda, \textit{Frobenius algebras and planar open
string topological field theories}, math.QA/0508349.

\bibitem[Mac]{Mac}  S. Mac Lane, \textit{Categories for the working
mathematician}, Springer, Berlin 1971.

\bibitem[GPS]{GPS}  R. Gordon, A. J. Power, R. Street, \textit{Coherence for
tricategories}, AMS, Providence 1995.

\bibitem[GS]{GS}  H. Grosse, K.-G. Schlesinger, \textit{On deformation
theory of quantum vertex algebras}, hep-th/0508225.

\bibitem[Sch]{Sch}  K.-G. Schlesinger, \textit{The universality question for
noncommutative quantum field theory}, in preparation.
\end{thebibliography}
\end{document}